\documentstyle[eqsecnum,aps,psfig,prl]{revtex}

\begin{document}
\twocolumn

\twocolumn[\hsize\textwidth\columnwidth\hsize\csname
@twocolumnfalse\endcsname

\preprint{AUTH-GR-00-1} \title{Nonlinear $r$-modes in Rapidly Rotating
Relativistic Stars} \author{N. Stergioulas${}^{(1)}$ and Jos{\'e}
A. Font${}^{(2)}$} \address{(1) Department of Physics, Aristotle
University of Thessaloniki, Thessaloniki 54006, Greece \\ (2)
Max-Planck-Institut f{\"u}r Astrophysik, Karl-Schwarzschild-Str. 1,
D-85740 Garching, Germany} \date{\today} \maketitle

\begin{abstract}
  The $r$-mode instability in rotating relativistic stars has been
  shown recently to have important astrophysical implications
  (including the emission of detectable gravitational radiation, the
  explanation of the initial spins of young neutron stars and the
  spin-distribution of millisecond pulsars and the explanation of one
  type of gamma-ray bursts), provided that $r$-modes are not saturated
  at low amplitudes by nonlinear effects or by dissipative mechanisms.
  Here, we present the first study of nonlinear $r$-modes in
  isentropic, rapidly rotating relativistic stars, via 3-D
  general-relativistic hydrodynamical evolutions. Our numerical
  simulations show that (1) on dynamical timescales, there is no
  strong nonlinear coupling of $r$-modes to other modes at amplitudes
  of order one -- unless nonlinear saturation occurs on longer
  timescales, the maximum $r$-mode amplitude is of order unity (i.e.,
  the velocity perturbation is of the same order as the rotational
  velocity at the equator). An absolute upper limit on the amplitude
  (relevant, perhaps, for the most rapidly rotating stars) is set by
  causality. (2) $r$-modes and inertial modes in isentropic stars are
  predominantly discrete modes and possible associated continuous
  parts were not identified in our simulations. (3) In addition, the
  kinematical drift associated with $r$-modes, recently found by
  Rezzolla, Lamb and Shapiro (2000), appears to be present in our
  simulations, but an unambiguous confirmation requires more precise
  initial data. We discuss the implications of our findings for the
  detectability of gravitational waves from the $r$-mode instability.


\end{abstract}


\narrowtext
\vskip2pc]

Considerable effort has been spent in the last two years, in
determining the properties of $r$-modes \cite{PP78} in rotating
compact stars, since the discovery \cite{A98,FM98} that these modes
are unstable to the emission of gravitational radiation. This is
motivated by the current understanding that the $r$-mode instability
may have several important astrophysical consequences: it provides an
explanation for the spin-down of rapidly rotating proto-neutron stars
to Crab-like spin-periods and for the spin-distribution of millisecond
pulsars and accreting neutron stars, while being a strong source of
detectable continuous gravitational radiation (see
\cite{S98,FL99,AK00} for reviews). In addition, if $r$-modes induce
differential rotation, then the interaction between them and the
magnetic field in neutron stars has been proposed as a gamma-ray burst
model \cite{S99} or as a mechanism for enhancing the star's toroidal
magnetic field, which could, in turn, limit the $r$-mode amplitude
\cite{RLS00}.

Still, several important uncertainties remain, the resolution of which
could significantly modify the above conclusions. In this {\it Letter}
we present nonlinear hydrodynamical simulations of $r$-modes in
rapidly rotating relativistic stars (in the relativistic Cowling
approximation \cite{MVS83}) and address the questions of the maximum
amplitude that $r$-modes can reach, the nature of their frequency
spectrum, and the existence of a kinematical differential drift,
associated with $r$-mode oscillations \cite{RLS00}. To our knowledge,
these are the first simulations of nonaxisymmetric oscillation modes
in rotating (Newtonian or relativistic) stars.

The maximum amplitude of unstable $r$-modes in a fluid star
(neglecting the magnetic field) and the precise mechanism by which it
is set, are currently unknown, but saturation of the amplitude is
thought to occur due to some form of nonlinear hydrodynamical
coupling.  An example of such a mechanism is the nonlinear coupling of
the unstable $r$-mode to other, stable, modes of
pulsation. Presumably, nonlinear saturation is set on a hydrodynamical
timescale, although it cannot be excluded that weak hydrodynamical
couplings saturate the $r$-mode amplitude on longer timescales (but
shorter than the growth timescale due to gravitational radiation
reaction).

For our study we have used a numerical code based on the 3-D
\verb+CACTUS+ code developed by the AEI-Potsdam/NCSA/Washington
University collaboration~\cite{Cactus}, in which we implemented the
3rd order Piecewise Parabolic Method (PPM) \cite{CW84} for the
hydrodynamics (see \cite{F00} for a recent review) and added initial
data for equilibrium and perturbed rapidly rotating relativistic
stars.  In \cite{FSK00} it was shown that the 3rd order PPM method is
suitable for long-term evolutions of rotating relativistic stars. The
equilibrium initial data are constructed using the \verb+RNS+ code
\cite{SF95}. We focus on a particular, representative, rapidly (and
uniformly) rotating model with gravitational mass $M=1.63M_\odot$,
equatorial {\it circumferential} radius $R=17.25$km and spin period
$P=1.26$ms (the gravitational mass and equatorial radius are larger
than for a corresponding nonrotating model, because of rotational
effects - the ratio of polar to equatorial coordinate radii is 0.7).
The star is rotating at 92\% of the mass-shedding (Kepler) limit (at
same central density) and we use the $N=1.0$ relativistic polytropic
equation of state (see \cite{S98}). Unless otherwise noted, we use
$116^3$ Cartesian grid-points, with an equidistant grid separation of
$\Delta x=\Delta y=\Delta z=0.31$km.

During the time-evolution, we only evolve the hydrodynamical
variables, keeping all spacetime variables fixed at their initial,
unperturbed, values (for small amplitude pulsations this is equivalent
to the Cowling approximation in linear perturbation theory; see
\cite{FSK00}). Since $r$-modes are basically fluid modes, we expect
that the adopted approximation is suitable for studying,
qualitatively, their basic properties.  The computational requirements
for a coupled spacetime and hydrodynamical evolution (for the same
long-term accuracy as reported here using the Cowling approximation),
by far exceed presently available supercomputing resources.

\vspace{-0.9cm}
\begin{figure}
 \centering \psfig{file=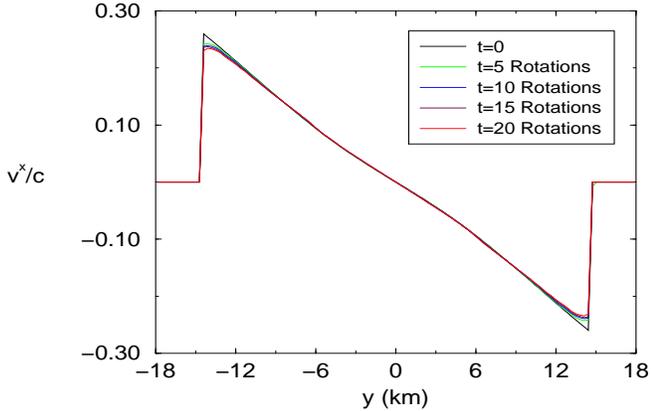,width=8.6cm,height=6.4cm}
\caption{Time-evolution of the rotational velocity profile for a
stationary equilibrium model, using the 3rd order PPM scheme. The star
remains stationary, even after 20 rotations. Larger deviations (due to
finite-differencing) occur only near the surface of the star.}
\label{fig_1}
\end{figure}
\vspace{-0.25cm}

Figure 1 clearly demonstrates the ability of the 3-D code to maintain
an unperturbed stationary equilibrium configuration (apart from the
influence of local truncation errors) in long-term evolutions. The
rotational velocity profile ($v^x$ along $y$-axis, where $v^i$ are the
contravariant components of the equilibrium 3-velocity, as measured by
a local zero-angular-momentum observer) remains nearly unchanged over
more than 20 rotational periods (all simulations presented here are
stopped due to computing time restrictions only and could be continued
for longer times).

In order to excite $r$-modes during the time-evolution, we perturb the
initial stationary model, by adding a specific perturbation $\delta v^i$ to
the equilibrium 3-velocity $v^i$. As there is no exact linear
eigenfunction available in the literature that would correspond to an
$l=m=2$ $r$-mode eigenfunction for rapidly rotating relativistic stars
in the Cowling approximation, we are using an approximate
eigenfunction, derived in spherical polar coordinates,
%
%
valid in the slow-rotation $O(\Omega)$ limit to the first
Post-Newtonian (1PN) order (see \cite{L99,LAF00} for details). 
%
%
%
We apply the usual transformation between the spherical polar and
Cartesian coordinate systems as the latter is used in the 3-D code.  
%
%
We note that, in the Newtonian limit, an amplitude of $\alpha=1.0$,
corresponds to a maximum velocity perturbation (i.e. the maximum value
of the velocity component $\delta v^\theta$ in the equatorial plane) equal to
roughly $1/3$ of the rotational velocity of the star at the equator. 
%
With
this eigenfunction (multiplied by a chosen initial amplitude $\alpha$, that
coincides, in the Newtonian limit, with the definition of the
amplitude in \cite{OLCSVA98}), we are able to excite mainly the
$l=m=2$ $r$-mode.  
%
%
Due to the approximate nature of the eigenfunction,
additional modes are also excited, primarily $m=2$ inertial modes.

\begin{figure}
  \centering
  \psfig{file=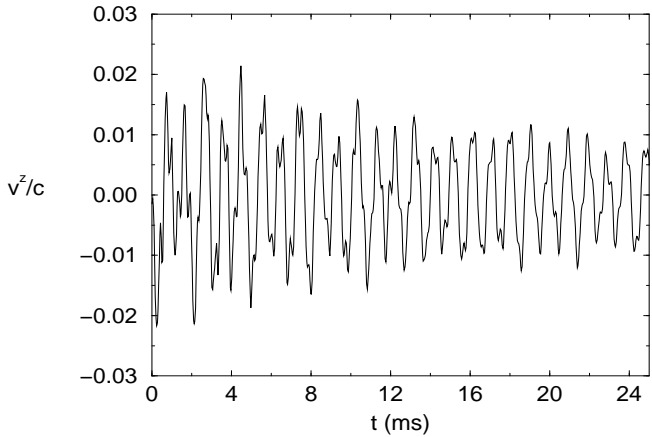,width=8.6cm,height=5.9cm}
  \caption{Evolution of the axial velocity in the equatorial plane for
  an amplitude of $\alpha=1.0$, at $r=0.75r_e$. The evolution is a
  superposition of (mainly) the $l=m=2$ $r$-mode and several inertial
  modes. The amplitude of the oscillation decreases due to numerical
  (finite-differencing) viscosity of the code. A beating between the
  $l=m=2$ $r$-mode and the $l_0=4,m=2$ inertial mode can also be
  seen.}
\label{fig_2}
\end{figure}

Figure \ref{fig_2} displays the evolution of the axial velocity in the
equatorial plane ($v^z$ along the $y$-axis) at a coordinate radius of
$r=0.75r_e$, where $r_e$ is the coordinate radius at the equator. It
is clear that the evolution is a superposition of several modes, and
that one mode, the $l=m=2$ $r$-mode, is the dominant component. The
amplitude in this evolution is $\alpha=1.0$. The perturbed star is evolved
for more than 25ms (26 $l=m=2$ $r$-mode periods), during which the
amplitude of the oscillation decreases due to numerical viscosity.
Even at amplitudes larger than 1.0, the evolution is still similar to
that in Figure \ref{fig_2}, with no sign of nonlinear saturation of
the $r$-mode amplitude on a dynamical timescale (only when $\alpha$
exceeds a value much larger than 1.0, nonlinear hydrodynamical
saturation sets in; however the precise determination of the maximum
saturation amplitude will require more accurate inital data and the
simultaneous dynamical evolution of the gravitational field).  Our
result implies, that, unless nonlinear saturation is set at timescales
much longer than the dynamical one, nonlinear hydrodynamical couplings
would not prohibit gravitational radiation reaction to drive unstable
$r$-modes to large amplitudes (of order one) before saturation sets
in.  An absolute upper limit on the $r$-mode amplitude (relevant,
perhaps, for the most rapidly rotating stars) is set by causality,
requiring $\sqrt{v_iv^i}<c$ (where $c$ is the speed of light), or
approximately, $\alpha_{\rm causal} \lesssim 3c/\Omega R$, where $\Omega$ is the angular
velocity of the equilibrium star (for slowly rotating stars, we expect
other upper limits to be more relevant).

\begin{figure}
  \centering
  \psfig{file=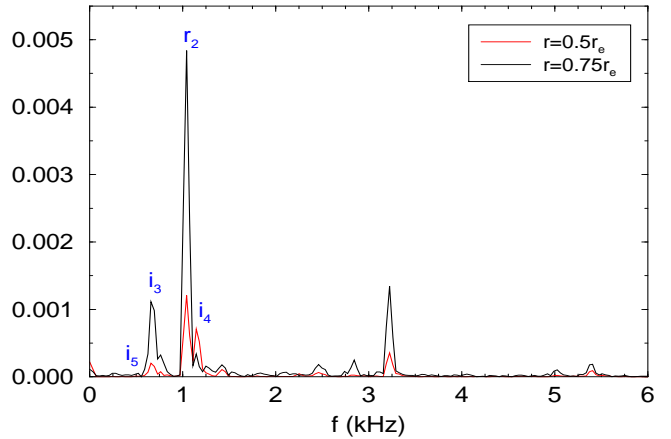,width=8.6cm,height=5.9cm}
  \caption{Fourier transform of the evolution shown in Figure
  {\ref{fig_2}}, showing the frequencies of the modes in the inertial
  frame. The frequencies of the modes are the same at different radii
  (discrete spectrum) and the ratio of the $l=m=2$ $r$-mode frequency
  to the frequencies of the identified inertial modes agrees with
  published results.}  \label{fig_3}
\end{figure}

A Fourier transform of the time-evolution shown in Figure \ref{fig_2},
as a function of the frequency in the inertial frame, shown in Figure
\ref{fig_3}, reveals that the initial data we are using, excite mainly
the $l=m=2$ $r$-mode ($r_2$), with a frequency of 1.03 kHz and, with
smaller amplitudes, several inertial modes ($i_3$, $i_4$, $i_5$) and
higher harmonics. The frequencies of the various modes are the same at
any given point inside the star (the specific data shown in the figure
correspond to radii $r= 0.5r_e$ and $r= 0.75r_e$), i.e. the evolution
consists of a sum of discrete modes and possible associated continuous
parts were not excited. This is in agreement with the conclusions
in \cite{LAF00}, regarding the $r$-mode spectrum in isentropic stars.
The possible existence of a continuous part in the frequency spectrum
of nonisentropic stars (see e.g. \cite{BK99}), will be examined in
future work.

\vspace{-0.15cm}
\vspace{-0.25cm}

In order to identify some of the peaks in the Fourier transform with
specific inertial modes, we compare the ratio of their frequencies
over the frequency of the $l=m=2$ $r$-mode to the corresponding ratios
derived from the normal mode linear eigenfrequencies, for a Newtonian
$N=1.0$ polytrope \cite{LF99}. This comparison shows good agreement
and allows us to identify, in Figure \ref{fig_3}, the $m=2$ inertial
modes with $l_o=3,4,5$ \cite{LF99}, having corresponding frequencies
$f=0.68, 1.16$ and $0.38$ kHz. In the Newtonian limit, one would
expect the above inertial modes to have frequencies $f=0.70, 1.14$ and
$0.37$ kHz, for an $l=m=2$ $r$-mode frequency of 1.03 kHz, which shows
that, even though the effects of relativity and rapid rotation on the
individual frequencies is not negligible, the ratios of the $r$-mode
frequency to the frequencies of the inertial modes remain close to
their Newtonian, slow-rotation values.

\begin{figure}
  \centering

  \psfig{file=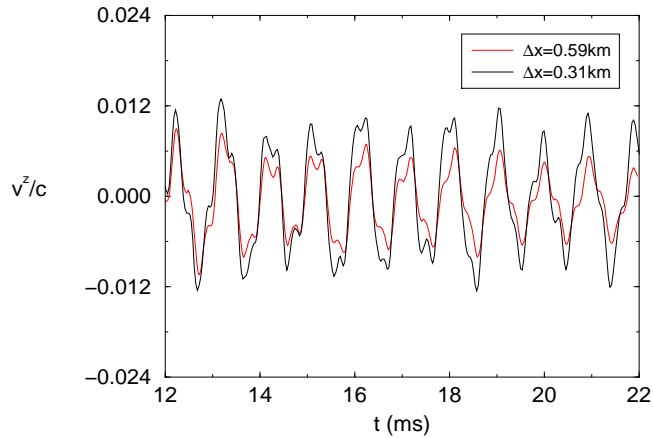,width=8.4cm,height=5.9cm}
  \caption{Late-time evolution of the axial velocity in the equatorial
  plane for $\alpha=1.0$, at $r=0.75r_e$, for two different
  grid-spacings.  The decrease in the oscillation amplitude scales as
  roughly first order with resolution (it is dominated by the
  first-order truncation error at the surface of the star).}
  \label{fig_4}
\end{figure}

We investigate the influence of the numerical viscosity of the code
(due to finite-differencing) on the evolution, by comparing evolutions
at different grid-spacings. Figure \ref{fig_4} shows the late-time
evolution of the axial velocity in the equatorial plane, at $r=
0.75r_e$ for grid-spacings of $\Delta x=0.59$km and $\Delta x=0.31$km. The
decrease in the amplitude of the oscillation (when compared to the
initial amplitude) scales as roughly first order with grid-spacing.
This indicates that the numerical viscosity damping the oscillations
is dominated by the truncation error at the surface of the star, which
is only first-order (compared to the second-order accuracy in the
interior; see related discussion in~\cite{FSK00}).

\vspace{-0.25cm}

In a recent paper \cite{RLS00} it is indicated that the hydrodynamical
oscillatory motion of an $r$-mode in a rotating star may be
accompanied by a differential drift (of kinematical origin). In the
equatorial plane, the drift velocity has opposite sign, compared to
the rotational velocity of the unperturbed star.  In a rotating star
with a poloidal magnetic field, this kinematical drift may wind up the
magnetic field lines and limit the effect of the $r$-mode instability
\cite{S99,RLS00}. The differential drift reported in \cite{RLS00} is
second order in the $r$-mode amplitude, but it is derived from the
first-order in amplitude definition of the $r$-mode velocity. Still,
in the restricted case of a spherical shell \cite{LU00}, the full
nonlinear hydrodynamical equations yield the same kinematical drift as
in \cite{RLS00}, independent of the amplitude of the oscillations.

Using our code we attempted to investigate the presence of such a
kinematical drift in the numerical evolutions. Our numerical results
do indicate that the perturbed star is rotating slower near the
surface (compared to the rotational velocity of the unperturbed star)
and that this drift scales roughly as $\alpha^2$ for amplitudes of
order $\alpha \sim 1.0$, although its magnitude is significantly
smaller than estimated in \cite{RLS00}.  We cannot test the presence
of the drift for amplitudes much less than $\alpha \sim 1.0$, since it
becomes smaller than numerical truncation errors.  Even at $\alpha
\sim 1.0$, it is not completely unambiguous that the drift we find is
the same as that predicted by \cite{RLS00}, as it could be due to the
fact that our initial data do not correspond exactly to a single
$l=m=2$ $r$-mode and thus also contain other modes and violate (to
some extent, which is still acceptable for the main purpose of this
work) the relativistic Hamiltonian and momentum constraints.  However,
the fact that our drift scales as $\alpha^2$, points to an association
with the results reported in \cite{RLS00}. Our preliminary findings on
the differential drift need to be confirmed by simulations with more
precise eigenfunctions and higher resolution, before a definite
conclusion can be drawn.

\vspace{-0.25cm}

In summary, our hydrodynamical evolutions of nonlinear $r$-modes show
that (1) nonlinear hydrodynamical couplings would not prohibit
$r$-modes in isentropic, rapidly rotating relativistic stars from
attaining a large amplitude, of order one, when driven unstable by
gravitational radiation reaction - it remains to be investigated
whether nonlinear saturation could set in at timescales longer than
the dynamical one, (2) $r$-modes and inertial modes, in isentropic
stars, are discrete and (3) a kinematical drift associated with
$r$-mode oscillations, appears to be present in our simulations,
although an unambiguous confirmation will require more precise initial
data.

Our finding that gravitational radiation could drive unstable
$r$-modes to a large amplitude, implies that $r$-modes can easily melt
the crust in newly-born neutron stars \cite{LOU00}, leaving the
initial conclusions about the $r$-mode instability being a strong
source of gravitational waves (see \cite{FL99,AK00}) essentially
unchanged. Our results also imply that $r$-modes could be excited to
large amplitudes during the violent formation of a proto-neutron star,
after a supernova core-collapse \cite{AJKS00}. Our finding that 
$r$-modes oscillate at a, predominatly, discrete frequency (at
least in isentropic stars) significantly simplifies the expected
gravitational wave signal compared to a signal emitted by a source
with a significant continuous part in its frequency spectrum. The
indications for the existence of the kinematical drift, call for a
more detailed consideration of the interaction between $r$-modes and
magnetic fields.

We will present details and tests of our numerical method, as well as
extensive higher-resolution results for various equations of state and
rotation rates, in a forthcoming paper \cite{SF00}. In future work, we
plan to study $r$-modes in nonisentropic stars and implement an
accelerated gravitational-radiation reaction force (see \cite{R99}).

It is a pleasure to acknowledge John Friedman for suggesting this
project as well as for helpful discussions and comments on the
manuscript. We also thank Nils Andersson, Curt Cutler, Kostas
Kokkotas, Yuri Levin and Luciano Rezzolla for discussions and comments
and the {\tt CACTUS} development team for making the code available.
The computations were carried out, in part, on a Cray T3E at the
Rechenzentrum Garching. We greatfully acknowledge computing time on an
Origin 2000, granted by the Albert-Einstein-Institute, Potsdam. This
research was supported, in part, by the European Union grant
HPRN-CT-2000-00137.

\end{document}